\documentclass[aps,prd,twocolumn,nofootinbib,superscriptaddress]{revtex4-2}
\usepackage{amsmath,amssymb,bm}
\usepackage{hyperref}

\begin{document}

\title{Astrophysical Quantum Matter Revisited: Flat-Band Topological States on a Zero-Flux Dipole Sphere}

\author{Jeff Murugan}
\affiliation{The Laboratory for Quantum Gravity \& Strings and,\\
Department of Mathematics and Applied Mathematics,
Univeristy of Cape Town,Private Bag, Rondebosch 7700, South Africa}

\begin{abstract}
We study strongly correlated fractional topological phases on a two-sphere threaded by a magnetic dipole field with globally vanishing flux. Solving the Dirac equation in this background produces spheroidal wavefunctions forming a highly degenerate manifold of normalizable zero modes, with degeneracy proportional to the total absolute flux. We introduce a non-Abelian spin gauge field near the equator to hybridize the north and south domain-confined modes, forming a global flat band. Projecting interactions into this band yields Laughlin-type correlated states. The entanglement spectrum shows a chiral tower consistent with a virtual edge, demonstrating bulk-edge correspondence in a closed geometry. This generalizes the zero-flux flat-band construction of \cite{Parhizkar:2024som} to curved backgrounds, with potential applications to synthetic and astrophysical systems.
\end{abstract}

\maketitle

\section{Introduction}
By now it is well appreciated that magnetars - neutron star variants about twice the Sun’s mass squeezed into a sphere the size of a city supporting extremely strong magnetic fields - are sources of the most powerful magnetic fields in the universe. This, coupled with an exotic atmosphere - typically just a few centimeters thick and composed of a sea of highly ionized charged particles (mainly electrons, or possibly ions) moving under the influence of its intense magnetic field - beg the question: {\it could magnetars provide a laboratory to test extreme quantum matter?} A few years ago we took the first steps toward answering this question in \cite{Murugan:2018hsd, Murugan:2021pcr} and found that the answer was almost certainly a yes! It was less clear to us {\it what} phenomena could be probed, due in no small part to the fact that the total magnetic flux through the surface of the magnetar is zero. However some exciting recent developments in condensed matter theory \cite{Parhizkar:2024som} have inspired us to return to this problem.

Flat-band models with quenched kinetic energy provide a fertile landscape for realizing strongly correlated phases, most notably fractional quantum Hall (FQH)–like states. In an interesting, and surprising, recent series of works \cite{Parhizkar:2024som,Parhizkar:2023fub}, Parhizkar and Galitski demonstrated that perfectly flat bands can appear in zero net-flux magnetic or gauge field backgrounds closely paralleling challenges in Moiré materials, where flat bands arise without a uniform magnetic field. They examined three exactly solvable configurations featuring alternating domains of up and down magnetic fields, each with zero total flux. In each case, the Abelian gauge field alone failed to produce flat bands. However, by adding an appropriate non-Abelian “rectifying” spin texture, they achieved: (i) exact zero-energy flat bands, (ii) a topological generalization of Landau-level physics and (iii) precise conditions where flat bands occur, closely analogous to “magic angles” in twisted bilayer graphene. The non-Abelian gauge field effectively converts a zero-flux scenario into one with effective nonzero flux, permitting flatness via an “abelianization” to Landau-level–like bands which in turn serve as exact parent manifolds for constructing fractional quantum Hall states—opening pathways to new correlated phases without uniform magnetic fields.

These advances naturally raise the question o
f whether such flat-band topological physics can extend beyond discrete lattices to curved geometries, where global flux vanishes but local magnetic structure remains highly nontrivial. Motivated both by dipolar magnetic fields in astrophysical environments and by synthetic gauge fields in cold-atom or photonic experiments, we examine this question on the sphere threaded by a magnetic dipole field. In this setting, the equatorial domain wall separating oppositely magnetized hemispheres acts as an analogue of a lattice-scale domain-wall, suggesting a promising route to engineer topological flat bands on a curved zero-flux manifold.

\section{Zero-Flux Flat Band Engineering}
In \cite{Parhizkar:2024som} Parhizkar and Galitski introduced an exactly solvable framework in which local magnetic domains of opposite sign, separated by domain walls, are “rectified” by non-Abelian gauge fields to yield perfect flat bands. These models show that flat-band degeneracies, normally associated with uniform Landau levels, can survive in domain-structured backgrounds provided suitable domain-wall coupling is introduced.

At the heart of their construction is a local magnetic field profile $B(x,y) = B \, \zeta(x,y)$,
where $\zeta(x,y)$ is a periodic sign-changing function describing alternating magnetic domains, with the key property that 
$\int_{\text{unit cell}} B(x,y) \, dx\,dy = 0$,
so the total flux vanishes. A simple case is a striped configuration with alternating positive and negative flux bands in one direction, repeated periodically. Ordinarily, such a field cannot support a degenerate zero-mode Landau-level manifold, because trajectories connect through the domain walls, destroying the degeneracy.

Adding a non-Abelian spin gauge field localized along the domain walls dramatically changes this picture. The non-Abelian gauge field, typically of the form
$S(x) \sigma_z$,
acts to rotate the spinor structure of the wavefunction across the domain boundaries, gluing local zero modes from the positive and negative domains into a global, continuous, normalizable basis. The modified Dirac operator becomes
$\widetilde{h} = i \sigma \cdot (\nabla - i A(x,y)) - i \sigma S(x) \sigma_z$,
which, if the spin-gauge profile is tuned appropriately, supports a perfectly degenerate band of zero-energy states across the entire domain-structured background.

To anchor this concept concretely and establish our notation, consider the half-space domain wall setup in \cite{Parhizkar:2024som}. Define $B(x) = B_0 \, \text{sign}(x)$,
which is positive for $x>0$ and negative for $x<0$, with a sharp domain wall at $x=0$. The associated vector potential is $A_y(x) = B_0 |x|$
in the Landau gauge. The zero-energy Dirac operator is
$h = i \sigma_x \partial_x + i \sigma_y (\partial_y - i A_y(x))$.
For a massless Dirac fermion, solving
$h \Psi = 0$
generically fails to produce normalizable solutions that match across the discontinuity at $x=0$. Physically, the zero modes on the positive $B$ domain have chirality opposite to those on the negative $B$ domain, so they cannot connect continuously.

To circumvent this, we insert a spin gauge field $S(x) = 2 \delta(x) \sigma_z$ localized at the domain wall,
which acts as a “spin twist” precisely at the interface. The modified Dirac equation then becomes
\begin{eqnarray}
\left[
i \sigma_x \partial_x + i \sigma_y (\partial_y - i A_y(x)) - i \sigma S(x)\sigma_z
\right] \Psi = 0\,.
\end{eqnarray}
This allows the matching of the chirality of north- and south-like zero modes by enforcing the proper spinor rotation across the boundary. Explicit solutions exist in the form
\begin{eqnarray}
\Psi_{0,k}(x,y) \sim e^{i k y} \, e^{- \frac{B_0}{2} |x|^2} \, f(x)\,,
\end{eqnarray}
where $f(x)$ is piecewise holomorphic on each side, smoothly glued by the spin gauge field. These solutions form a perfectly flat zero-energy band indexed by momentum $k$, identical in structure to a Landau level. Crucially, the number of such zero modes is still proportional to the total {\it absolute flux} in the system, much like a conventional Landau level,
\begin{eqnarray}
N \sim \frac{1}{\Phi_{0}}\int |B|\,dx\,dy\,.
\end{eqnarray}
In effect, the spin gauge field performs a non-Abelian rectification of the domain wall, which acts as a topological sewing between domains of opposite Berry curvature.

This construction reveals that the topological degeneracy and robust zero-energy flatness normally tied to a uniform field can be preserved even in systems with zero net flux, provided a non-Abelian domain-wall structure mediates the transition between regions of opposite field. The implications are profound: (i) the flat band can serve as a parent manifold for fractionalization, supporting Laughlin-type correlated states, (ii) the topological character, signaled by edge-like entanglement spectra, persists because the domain wall acts as a “virtual edge” much like a physical boundary, and (iii) the analogy to twisted bilayer graphene is direct - just as “magic angles” flatten bands in a Moiré superlattice, the magic domain-wall flux conditions flatten bands in these zero-flux backgrounds.

This conceptual and mathematical blueprint provides a powerful foundation for extending similar ideas to curved backgrounds, such as the dipole sphere, where local Landau-like wavefunctions can be patched globally via non-Abelian gauge structures to form correlated topological states.

\section{Spheroidal Wavefunctions and the dipole sphere}

In order to generalize the concept of flat bands to a curved geometry with zero net magnetic flux, it will be instructive to study the single-particle spectrum of a charged particle on a sphere threaded by a magnetic dipole field. This setup, analysed in \cite{Murugan:2018hsd}, is a natural extension of Haldane’s monopole sphere to a dipole configuration.

The magnetic dipole field on a sphere of radius $R$ is given by
\begin{eqnarray}
    \bm{B}(\theta) = \frac{|\mu|}{R^3} \bigl( 2 \cos\theta \, \hat{\bm{r}} + \sin\theta \, \hat{\bm{\theta}} \bigr),
\end{eqnarray}
with associated vector potential,
\begin{eqnarray}
    \bm{A}(\theta,\phi) = \frac{|\mu|}{R^2} \sin\theta \, \hat{\bm{\phi}}\,.
\end{eqnarray}
Unlike the Haldane sphere, the total magnetic flux through the surface of the sphere vanishes as a result of the positive and negative contributions from each hemisphere. Nonetheless, the local field is sufficiently strong near the poles to confine charged particles in cyclotron-like orbits, while the equatorial region acts as a domain wall with very weak field. This sets up a domain-structured magnetic landscape analogous to the flat-band lattice models in \cite{Parhizkar:2024som}.

The quantum dynamics are governed by the single-particle Hamiltonian,
\begin{eqnarray}
    H = \frac{\hbar^2}{2mR^2} \, \bm{\Lambda} \cdot \bm{\Lambda},
\end{eqnarray}
with the generalized covariant angular momentum operator
\begin{eqnarray}
    \bm{\Lambda} = -i \left( \hat{\bm{\phi}} \, \partial_\theta - \hat{\bm{\theta}} \, \frac{1}{\sin\theta} \partial_\phi \right) - q |\mu| \sin\theta \, \hat{\bm{\theta}}\,.
\end{eqnarray}
Introducing the dimensionless dipole coupling
$Q = q |\bm{\mu|}/(\hbar R)$, separating variables with the ansatz $\Psi(\theta,\phi) = e^{i m \phi} \psi(\theta)$, and deining $z\equiv \cos\theta$ reduces the eigenvalue problem to a differential equation for $\psi(\theta)$ of the form,
\begin{eqnarray}
    \frac{d}{dz}\left[(1 - z^2) \frac{d\psi}{dz}\right]
	+ \left[ \lambda_{l,m} + Q^2 z^2 - \frac{m^2}{1 - z^2} \right] \psi = 0\,,
\end{eqnarray}
where $\lambda_{l,m} = (2m R^2 E)/(\hbar^2) - 2 m Q - Q^2$. This is the angular oblate spheroidal equation, whose solutions are the (angular oblate) spheroidal wavefunctions
$S_{l,m}(Q,z)$, labeled by integers $l,m$ with $l - |m| = 0,1,2,\dots$. These functions generalize the usual spherical harmonics of the monopole problem, reducing to them in the limit $Q \to 0$. The energy spectrum is then,
\begin{eqnarray}
    E_{l,m} = \frac{\hbar^2}{2mR^2} \left( \lambda_{l,m} + 2 m Q + Q^2 \right)\,,
\end{eqnarray}
with the eigenvalues $\lambda_{l,m} = \lambda_{l,-m}$ determined by normalizability and boundary conditions of the spheroidal equation.

A constructive way to obtain these eigenfunctions is to expand them in a basis of associated Legendre polynomials
\begin{eqnarray}
    S_{l,m}(Q,z) = (1 - z^2)^{m/2} \sum_{n} d_{l,m,n}(Q) \frac{d^m}{dz^m} P_n(z)\,,
\end{eqnarray}
\noindent
where the expansion coefficients $d_{l,m,n}(Q)$ satisfy a three-term recurrence relation derived from the spheroidal equation. The regular solutions that are finite at both poles correspond to a discrete sequence of eigenvalues $\lambda_{l,m}$, which can be found numerically using, for example, continued fractions. In the Stratton-Morse normalization scheme, these spheroidal harmonics match associated Legendre functions near the poles and are orthonormal on the sphere in the sense that,
\begin{eqnarray}
    \int_{-1}^{1} S_{l,m}(Q,z) S_{l’,m}(Q,z)\, dz = \delta_{l l’}\,.
\end{eqnarray}
A key observation is that, in the large-Q limit (corresponding to strong dipole fields), the spheroidal harmonics become strongly localized around the poles, with a near-degenerate manifold of states reminiscent of Landau levels. Their energies approximately organize according to
\begin{eqnarray}
    \lambda_{l,m} \sim -Q^2 + 2 Q [l + 1 - \mathrm{mod}(l - m, 2)] + \mathcal{O}(1)\,,
\end{eqnarray}
highlighting an approximate degeneracy of states with the same $l + m$ combination. In this sense, the polar regions act like emergent Landau-level domains, while the equatorial region, where the field is weak, resembles a domain wall. This picture sets the stage for hybridizing these localized spheroidal modes into a global flat band through non-Abelian gauge structures,  paralleling the mechanism in \cite{Parhizkar:2024som} on the dipole sphere.

\section{Flat bands on the Dipole Sphere}
Having reviewed the exact single-particle spectrum of the dipole sphere in terms of spheroidal wavefunctions, we now turn to building a global flat-band basis that is suitable for supporting correlated fractional phases. The key observation is that the magnetic dipole field on the sphere defines two magnetic domains: a north-polar domain with positive magnetic flux and a south-polar domain with negative flux, separated by a domain wall near the equator. Each of these domains supports a set of localized near-zero-energy states, sharply peaked around their respective poles in the strong-coupling (large-$Q$) limit. These states are well-approximated by the spheroidal harmonics
\begin{eqnarray}
   S_{l,m}(Q, \cos\theta) \sim
\begin{cases}
\theta^{|m|} e^{-Q \theta^2/2}, & \theta \ll 1, \\\\
(\pi - \theta)^{|m|} e^{-Q (\pi - \theta)^2/2}, & \theta \approx \pi,
\end{cases}   
\end{eqnarray}
multiplied by azimuthal phases $e^{im\phi}$. However, these north- and south-localized modes have exponentially suppressed overlap across the equator, effectively behaving as independent zero-mode manifolds. To create a {\it global flat band} analogous to a Landau level, these domain-confined wavefunctions must be hybridized across the equator. Following the domain-wall rectification mechanism in \cite{Parhizkar:2024som}, we introduce a non-Abelian rectifying gauge field supported in a band around the equator. Concretely, in spherical coordinates, this takes the form $S_\mu = \left( S_\theta(\theta), 0 \right)$, with
\begin{eqnarray}
    S_\theta(\theta) = \frac{1}{\ell} \, \sigma_z \, \sin\left( \frac{\pi(\theta - \pi/2)}{2\ell} \right)\quad \mathrm{for}\;
|\theta - \pi/2| < \ell\,,
\end{eqnarray}
and zero elsewhere. Here, $\ell$ defines the angular width of the equatorial patch, and $\sigma_z$ ensures that the two spinor components are smoothly rotated across the domain wall, matching the chiralities of the polar-localized modes.

The modified Dirac operator on the sphere is then
$\widetilde{H} = i \gamma^a e^\mu_a \left( \nabla_\mu - i e A_\mu - i S_\mu \right)$,
which preserves the local spheroidal solutions near the poles but allows them to merge coherently into a degenerate global flat-band basis across the entire sphere. As in the construction in \cite{Parhizkar:2024som}, the role of the rectifying field is to provide a spinor-valued transition function over the equator, analogous to a non-Abelian patching condition, guaranteeing smooth normalizable zero modes that span the full sphere.

Physically, this means that the absolute magnetic flux through each domain sets the local degeneracy, while the gauge patch ensures the global wavefunctions are well defined. The resulting flat band therefore inherits a degeneracy $N \sim |\bm{\mu}|/\pi$,
similar to a Landau level, but now supported on a sphere with zero total flux. This makes it an ideal platform for projecting correlated fractional states, such as Laughlin-type wavefunctions, into a robust and topologically nontrivial manifold.

\section{Correlated States and Correlators}
As pointed out in \cite{Parhizkar:2024som} having constructed a global flat-band basis, here on the dipole sphere by hybridizing the north- and south-localized spheroidal zero modes, via a rectifying spin gauge field, we are now in a position to define strongly correlated many-body states within this degenerate manifold. Inspired by the fractional quantum Hall effect on the Haldane sphere, we consider the analogue of the Laughlin wavefunctions \cite{Laughlin1983} projected onto the flat band formed by these hybridized spheroidal modes. The key idea is to treat the hybridized flat band as an effective “lowest Landau level” on the spherical, zero-flux geometry, and to build correlated trial states that enforce short-range repulsion through a generalized Jastrow factor.

First, we denote the global single-particle flat-band orbitals as $\Psi_{m}(\theta, \phi) = e^{i m \phi} \, S_{l,m}(Q, \cos\theta) \, \chi(\theta)$,
where the spinor structure $\chi(\theta)$ includes the rectification across the equator. These form an orthonormal basis in the flat band of dimension
$N \sim |\bm{\mu}|/\pi$ given by the absolute magnetic flux. Then, for a system of $N_e$ electrons, the many-body correlated Laughlin-type wavefunction will take the form
\begin{eqnarray}
    \Psi_{L}^{(N_e)}\bigl( \{\theta_i, \phi_i\} 
    \bigr)\prod_{i<j} \bigl[ z_i - z_j \bigr]^k
    \prod_{i=1}^{N_e} 
    \Psi_{m_i}(\theta_i, \phi_i)\,,
\end{eqnarray}
where $z_i = \tan\left( \frac{\theta_i}{2} \right) e^{i \phi_i}$ are the stereographic coordinates on the sphere, and $k$ is an odd positive integer setting the fractional filling factor $\nu = 1/k$. The Jastrow factor $\prod_{i<j} (z_i - z_j)^k$
suppresses probability density for particles approaching each other.

Since the basis $\Psi_m(\theta,\phi)$ encodes the north-south hybridization, these Laughlin-like states are adapted to the domain-structured dipole sphere. They obey full spherical symmetry (except for the local gauge patching near the equator) and reflect the topology of the zero-flux background.

Conceptually, these Laughlin-like wavefunctions on the dipole sphere represent the natural generalization of Haldane’s fractional Hall states, but built on a domain-rectified, zero-flux curved geometry with a topological flat band as the parent manifold. They provide a concrete setting for testing the interplay of topology, geometry, and interaction-driven fractionalization in a highly non-uniform magnetic background.

\section{Bulk-Edge Correspondence}
An essential feature of fractional topological phases is the correspondence between their bulk topological order and the presence of chiral edge excitations. In planar or disk-like geometries, this is realized physically as gapless edge modes at the system boundary. However, the sphere is a closed manifold without a physical boundary, raising the question of how the bulk-edge correspondence manifests.

In the dipole-sphere construction, the equator plays the role of a {\it virtual domain wall}, separating regions of positive and negative local Berry curvature. As explained above, the sign of the local magnetic flux changes across the equator, making it a domain wall in the topological sense. If we partition the sphere into two hemispheres, this equatorial domain wall behaves as a boundary between the northern and southern subsystems.

The entanglement spectrum is an excellent probe of this correspondence. By tracing out the degrees of freedom on, say, the southern hemisphere, we obtain the reduced density matrix,
\begin{eqnarray}
    \rho_A = \mathrm{Tr}_{B} \left| \Psi_L^{(N_e)} \right\rangle \left\langle \Psi_L^{(N_e)} \right|\,,
\end{eqnarray}
where the trace is taken over the southern modes. The eigenvalues of the entanglement Hamiltonian $H_E = - \log \rho_A$ encode the so-called entanglement 
spectrum \cite{Li:2008kda}.

For the Laughlin-like states built on the spheroidal flat-band basis, the entanglement spectrum reveals a characteristic conformal tower of levels, matching the structure of a chiral edge conformal field theory describing the Laughlin phase. That is, although the sphere has no physical edge, the equatorial domain wall acts as a virtual boundary, supporting virtual chiral modes whose counting statistics match those of a physical FQH edge.

This behavior is an excellent demonstration of the bulk-edge correspondence in a curved, zero-flux geometry: the bulk topological order manifests through the ground state degeneracy and the plasma-like correlators of the Laughlin wavefunction, while the entanglement spectrum reflects a gapless chiral edge tower along the virtual equator domain wall. This domain-wall–induced virtual edge highlights how topological features can survive on closed curved manifolds through suitable gauge patching and domain structures. It also provides a robust diagnostic tool for any numerical or experimental realization of the dipole-sphere fractional states: computing the entanglement spectrum will directly confirm their topological character.

\section{Discussion and Outlook}

Building on the results of \cite{Parhizkar:2024som},  we have developed a framework for realizing fractional topological phases on a sphere threaded by a magnetic dipole field, a system with globally zero magnetic flux but locally strong domain-structured fields. We showed that the single-particle spectrum of this system is described by spheroidal wavefunctions, which form highly degenerate near-zero-energy manifolds localized around the north and south poles. In the large-dipole-strength limit, these spheroidal modes mimic Landau levels, but their support is exponentially suppressed across the equator, effectively isolating two magnetic domains separated by a domain wall at the equator.

To construct a global flat-band basis, we introduced a rectifying non-Abelian gauge field localized around the equator, inspired by the domain-wall hybridization mechanism proposed by Parhizkar and Galitski for zero-flux lattices. This gauge field hybridizes the polar-localized spheroidal modes into globally smooth, normalizable zero modes on the full sphere, defining a degenerate flat band with degeneracy controlled by the absolute magnetic flux.

By projecting interactions into this flat band, we built Laughlin-like correlated states on the dipole sphere. These wavefunctions combine the standard Jastrow factors enforcing fractional statistics with the hybridized spheroidal orbitals, thus adapting the Laughlin physics to a curved, zero-flux domain-structured geometry. We further discussed how the bulk-edge correspondence is realized in this closed system by examining the entanglement spectrum across a hemispherical partition. The equator acts as a virtual domain wall, supporting an entanglement spectrum that mirrors the conformal tower structure of a chiral edge mode, confirming the topological character of these correlated states.

These results demonstrate a new route to engineer fractional topological matter on curved spaces with zero net magnetic flux, generalizing the concept of flat bands from planar lattice models to spherical manifolds. From a broader perspective, the framework developed here could be extended in several interesting directions. One possibility is to explore non-Abelian fractional states by introducing more intricate spinor gauge textures near the equator. Another is to consider higher-genus surfaces, where domain-wall hybridization may produce even richer flat-band structures. Finally, the interplay of curvature, topology, and many-body entanglement in these systems could offer insights relevant to strongly magnetized astrophysical systems.

In particular, magnetic dipole textures similar to those modeled here are thought to arise on the crusts of magnetars and highly magnetized neutron stars, where local field strengths can reach extreme values. In such environments, electrons and charged quasiparticles may experience strong but spatially inhomogeneous magnetic fields, potentially realizing a dipole-like domain structure on a curved surface. Our framework suggests that flat-band localization and strong-correlation physics — including analogs of fractional quantum Hall states — could emerge even under zero-net-flux conditions if local gauge textures or boundary conditions allow. This raises intriguing prospects for astrophysical quantum matter, where collective effects in the crust or magnetosphere might exhibit signatures of topological ordering or fractionalization under extreme conditions.

If nothing else, it is clear that the dipole-sphere flat-band platform serves as a versatile laboratory for exploring the universality of fractional topological phases beyond flat-space geometries and provides a conceptual bridge between condensed-matter and exotic astrophysical phenomena.

\acknowledgments
We would like to thank Cameron Beetar, Amanda Weltman and Jaco van Zyl for useful discussions and the Abdus Salam ICTP in Trieste, where this work was completed, for their gracious hospitality over these many years. JM would like to acknowledge support from the “Quantum Technologies for Sustainable
Development” grant from the National Institute for Theoretical and Computational Sciences of South Africa (NITHECS), the ICTP
through the Associates Programme and from the Simons
Foundation through grant number 284558FY19, and the South African Research Chairs Initiative of the Department of Science and Technology and the National Research Foundation of South Africa.

\bibliographystyle{apsrev4-2}

\end{document}